\documentclass{PoS}

\title{Equation of state for two-flavor QCD with an improved Wilson 
quark action at non-zero chemical potential}

\ShortTitle{Equation of state for two-flavor QCD with an improved Wilson 
quark action at non-zero chemical potential}

\author{\speaker{S. Ejiri}, T. Hatsuda, N. Ishii, Y. Maezawa and N. Ukita\\
        Department of Physics, The University of Tokyo, 
        Tokyo 113-0033, Japan\\
        E-mail: \email{ejiri@nt.phys.s.u-tokyo.ac.jp}}
\author{S. Aoki and K. Kanaya\\
        Institute of Physics, University of Tsukuba, 
        Tsukuba, Ibaraki 305-8571, Japan}

\abstract{
  The QCD thermodynamics on the lattice  
  provides fundamental theoretical grounds to analyze the 
  various experimental data in relativistic heavy ion collisions.
  So far, most of the numerical simulations on the lattice 
  have been performed by using the 
  staggered-type fermion actions. Therefore it 
  is important to carry out studies using  different 
  fermion formulations to test the uncertainties
  of the lattice QCD results. For this purpose, 
  we perform systematic simulations of two-flavor QCD with an improved 
  Wilson quark action to investigate the equation of state.
  We report the current status of our project and show the preliminary 
  results of the Taylor expansion coefficients of the thermodynamic 
  grand partition function in terms of chemical potential.}

\FullConference{XXIVth International Symposium on Lattice Field Theory\\
                July 23-28, 2006\\
                Tucson, Arizona, USA}

\begin{document}

\section{Introduction}

Numerical simulation of lattice QCD at non-zero temperature $(T)$ and 
quark chemical potential $(\mu_q)$ is an essential tool for 
quantitative understanding of the QCD phase transition.
So far, most of the studies have been performed using the staggered-type quark 
actions with the fourth-root trick of the quark determinant. 
To test the uncertainties of the lattice QCD results due to 
different fermion formulations and to obtain
basic information to analyze the experimental data in
relativistic heavy ion collisions, 
systematic studies of the QCD thermodynamics using a 
Wilson-type quark action are called for. 
Such a study at $T\neq 0$ and $\mu_q =0$ has been initiated six years ago
using the Iwasaki (RG) improved gauge action and the $N_f=2$ clover improved 
Wilson quark action by the CP-PACS Collaboration \cite{PSt,EoS}.  
The phase structure, the transition temperature and the equation of 
state have been investigated in detail, and also the crossover scaling 
in the region near the chiral phase transition point has been tested. 
It is confirmed that a subtracted chiral condensate satisfies the 
scaling behavior with the critical exponents and scaling function of 
the 3-dimensitonal O(4) spin model, suggesting the chiral phase transition 
is in the same universality class as the O(4) spin model. 
Moreover calculations of various physical quantities at $T=0$ such as the light 
hadron masses have been carried out using the same action \cite{LHS}.

Since there are numbers of technical progresses in treating 
the system with finite baryon density in the past six years,
it may be worth while to revisit the QCD thermodynamics with
Wilson-type quarks, and to study especially the effect of non-zero 
baryon density.
In this report, we will highlight the fluctuations at non-zero 
temperature and density among various topics we are studying by
Wilson quarks.
The existence of the endpoint of the first order phase transition 
in the $(T, \mu_q)$ plane is suggested in phenomenological studies 
and has attracted much attentions both in theories and experiments.
Among others, an interesting result has been reported in numerical 
simulations of the quark number susceptibility (the second 
derivative of the thermodynamic grand canonical potential $\Omega$) 
in the framework of the Taylor expansion with respect to $\mu_q/T$
by the Bielefeld-Swansea Collaboration \cite{BS03}.
By calculating the Taylor expansion coefficients 
of $\Omega$ up to $O[(\mu_q/T)^6]$ using the p4-improved 
staggered fermions, 
they found that the quark number fluctuation increases rapidly 
as $\mu_q$ increases in the region near the transition temperature. 
This suggests indirectly the existence of the critical point in the 
$(T,\mu_q)$ plane. 
It is particularly important to confirm whether the same result 
is obtained using the Wilson fermions  which does not resort to 
the fourth-root trick of the quark determinant.
Because the odd derivatives of $\Omega$ vanishes and the second derivative 
is the susceptibility at $\mu_q=0$, the forth derivative is the leading 
term necessary to investigate the $\mu_q$-dependence of the susceptibility. 
We calculate the second and fourth derivatives of $\Omega$ with 
respect to $\mu_q$, and discuss its behavior at finite $\mu_q$.

\section{Simulations}

\begin{figure}[t]
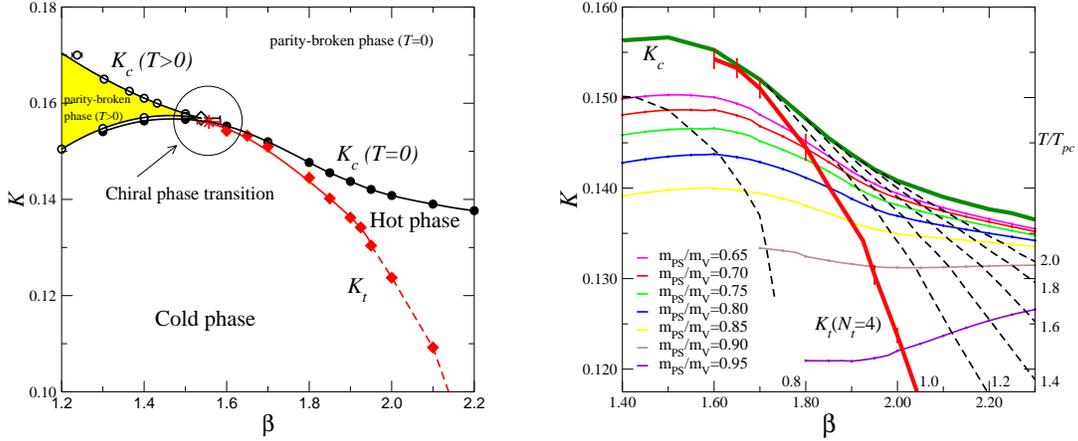

\begin{center}
\includegraphics[width=2.7in]{phasedia.eps}
\hskip 0.5cm
\includegraphics[width=2.7in]{bkconst.eps}
\vskip -0.2cm
\caption{Left: Phase structure of QCD with Wilson-type quarks. 
Right: Lines of constant $m_{PS}/m_V$ and lines of constant 
$T/T_{pc}$.}
\label{fig1}
\end{center}
\vskip -0.3cm
\end{figure} 

We perform simulations for $N_f=2$ at $m_{PS} / m_V = 0.80$.
We adopt the Iwasaki (RG) improved gauge action and the clover improved 
Wilson fermion action: 
\begin{eqnarray}
{\cal Z}(\beta, K, \mu) &=& \int 
{\cal D}U (\det M)^{N_f} e^{-S_g},
\hspace{5mm}
 S_g = -\beta \left\{ \sum_{x,\, \mu > \nu} 
      c_{0} W_{\mu \nu}^{1 \times 1}(x) 
    + \sum_{x,\, \mu, \nu} c_{1} W_{\mu \nu}^{1 \times 2}(x) \right\},
\nonumber \\
  M_{x,y} &=& \delta_{x,y}
  - \delta_{x,y} c_{SW} K \sum_{\mu > \nu} \sigma_{\mu \nu} F_{\mu \nu} 
    -K \sum_{i} \left[(1-\gamma_i) U_i(x)~\delta_{x+\hat{i},y}
    +(1+\gamma_i){U_i}^\dagger(x-\hat{i})~\delta_{x-\hat{i},y}\right]
     \nonumber \\
  && -K \left[e^{\mu} (1-\gamma_4) U_4(x)~\delta_{x+\hat{4},y}
    -e^{-\mu} (1+\gamma_4) {U_4}^\dagger(x-\hat{4})~\delta_{x-\hat{4},y}
    \right],
  \label{eq:fermact}
\end{eqnarray}
where $W_{\mu \nu}^{1 \times 1}(x)$ and $W_{\mu \nu}^{1 \times 2}(x)$ are 
$1 \times 1$ and $1 \times 2$ Wilson loops, 
$F_{\mu \nu}=(f_{\mu \nu} - f_{\mu \nu}^\dagger)/(8i)$, 
$f_{\mu \nu}$ is the standard clover-shaped combination of gauge links, 
$\beta=6/g^2$, $c_1=-0.331$, $c_0=1-8c_1$, 
$c_{SW}=(1-0.8412 \beta^{-1})^{-3/4}$, and $\mu \equiv \mu_qa$.

The phase structure of QCD at $\mu_q=0$ with this action has been 
investigated in Ref.~\cite{PSt}.
The black line $(K_c)$ in Fig.~\ref{fig1} (left) is the chiral limit, 
on which the pion mass vanishes at zero temperature.
Above this line, a parity-flavor symmetry is spontaneously broken. 
Numerical simulations are performed in the normal phase below $K_c$. 
At finite temperature, the parity-flavor broken phase becomes narrow, 
that is the colored region in the upper left of Fig.~\ref{fig1} (left) 
for $N_t=4$. 
On the boundary of this phase, pion mass vanishes. 
The red line $(K_t)$ is the finite temperature pseudo-critical line 
for $N_t=4$, separating the cold phase at small $\beta$ and the 
hot phase at large $\beta$. 

The relation between the simulation parameters $(\beta, K)$ and 
the physical parameters is shown in Fig.~\ref{fig1} (right). 
We determine the lines of constant physics (LCP's) by the mass ratio of 
pseudo-scalar meson and vector meson $m_{PS}/m_{V}$, 
interpolating the data of $m_{PS}$ and $m_{V}$ at 
$T=0$ in Refs.~\cite{PSt,EoS,LHS}.
The temperature is estimated by the vector meson mass $m_V a$ using 
$T/m_V=1/(N_t m_V a)$, and normalized by $T_{pc}/m_V$, where $T_{pc}$ 
is the temperature at the pseudo-critical point on each LCP 
(see Sec.IV in Ref.~\cite{EoS}).
Colored lines in Fig.~\ref{fig1} (right) are the lines of constant 
$m_{PS}/m_V$ (LCP's) and dashed lines are the lines of constant $T/T_{pc}$.
$(T/T_{pc}=0.8-2.0.)$
In this study, we generate 500 - 600 configurations (5000 - 6000 trajectories) 
for each simulation point. 
The lattice size is $N_s^3 \times N_t = 16^3 \times 4$.
The details of our simulations are given in Ref.~\cite{ukita}.

\section{Taylor expansion of the grand canonical potential}

QCD at finite density is known to have a serious problem called 
``the sign problem''.
To avoid this problem, we perform a Taylor expansion of physical 
quantities in terms of $\mu_q$ 
around $\mu_q=0$ and calculate the expansion coefficients, 
i.e. derivatives of the physical quantities, by numerical simulations 
at $\mu_q=0$, so that this calculation is free from the sign problem.
The calculations of the derivatives are basic measurements in the study 
of QCD thermodynamics, since most of thermodynamic quantities are given 
by the derivatives of partition function ${\cal Z}(T,\mu_u,\mu_d)$, e.g.,
pressure $p$ and quark number density $n$ are defined by
\begin{eqnarray}
\frac{p}{T^4} = \frac{1}{VT^3} \ln {\cal Z} \equiv \Omega,
\hspace{8mm}
\label{eq:qnd}
\frac{n_{u(d)}}{T^3} = 
\frac{1}{VT^3} \frac{\partial \ln {\cal Z}}{\partial (\mu_{u(d)}/T)} 
= \frac{\partial (p/T^4)}{\partial (\mu_{u(d)}/T)}, 
\end{eqnarray}
where $\mu_{u(d)}$ is the chemical potential for the u(d) quark.
Quark number susceptibility $(\chi_q)$ and
isospin susceptibility $(\chi_I)$ are given by
\begin{eqnarray}
\frac{\chi_q}{T^2} 
= \left( \frac{\partial}{\partial (\mu_u/T)} 
+ \frac{\partial}{\partial (\mu_d/T)} \right) 
\frac{n_u + n_d}{T^3}, \hspace{5mm} 
\frac{\chi_I}{T^2} 
= \left( \frac{\partial}{\partial (\mu_u/T)} 
- \frac{\partial}{\partial (\mu_d/T)} \right) 
\frac{n_u - n_d}{T^3}. 
\end{eqnarray}
Moreover the chiral condensate is defined by the derivative of 
$\ln {\cal Z}$ with respect to the quark mass.

We define the Taylor expansion coefficients of the pressure 
$p(T, \mu_q)$ for the case $\mu_u=\mu_d \equiv \mu_q$ as 
\begin{equation}
\frac{p}{T^4} =
\sum_{n=0}^\infty c_n(T) \left(\frac{\mu_q}{T}\right)^n,
\label{eq:p}
\hspace{8mm}
c_n (T)= 
\frac{1}{n!} \frac{N_t^{3}}{N_s^3} \left.
\frac{\partial^n \ln{\cal Z}}{\partial(\mu_q/T)^n} \right|_{\mu_q=0}.
\label{eq:cn}
\end{equation}
The quark number and isospin susceptibilities for 
$\mu_u=\mu_d \equiv \mu_q$ are given by
\begin{eqnarray}
\frac{\chi_q(T,\mu_q)}{T^2}
=2c_2+12c_4\left(\frac{\mu_q}{T}\right)^2+\cdots ,
\hspace{5mm}
\frac{\chi_I(T,\mu_q)}{T^2}
=2c^I_2+12c^I_4\left(\frac{\mu_q}{T}\right)^2+\cdots ,
\end{eqnarray}
where
\begin{equation}
c^I_n= \left. \frac{1}{n!}\frac{N_t^{3}}{N_s^3}
\frac{\partial^n \ln {\cal Z}(T,\mu_q+\mu_I,\mu_q-\mu_I)}
{\partial(\mu_I/T)^2 \partial(\mu_q/T)^{n-2}}
\right|_{\mu_q=0,\mu_I=0}, 
\hspace{5mm}
\mu_I=\frac{\mu_u - \mu_d}{2}.
\end{equation}

The explicit forms of the coefficients are
\begin{eqnarray}
&& \hspace{-13mm}
c_2 = \frac{N_t}{2N_s^3} {\cal A}_2 ,
\hspace{3mm}
c_4 = \frac{1}{4! N_s^3 N_t} ({\cal A}_4 -3 {\cal A}_2^2) , 
\hspace{3mm}
c_2^I = \frac{N_t}{2N_s^3} {\cal B}_2 ,
\hspace{3mm}
c_4^I = \frac{1}{4! N_s^3 N_t} ({\cal B}_4 - {\cal B}_2 {\cal A}_2) , \\
{\cal A}_2 &=&
\left\langle {\cal D}_2 \right\rangle 
+\left\langle {\cal D}_1^2 \right\rangle, 
\hspace{5mm}
{\cal A}_4 =
\left\langle {\cal D}_4 \right\rangle 
+4\left\langle {\cal D}_3 {\cal D}_1 \right\rangle 
+3\left\langle {\cal D}_2^2 \right\rangle 
+6\left\langle {\cal D}_2 {\cal D}_1^2 \right\rangle 
+\left\langle {\cal D}_1^4 \right\rangle , \nonumber \\
{\cal B}_2 &=&
\left\langle {\cal D}_2 \right\rangle ,
\hspace{5mm}
{\cal B}_4 =
\left\langle {\cal D}_4 \right\rangle 
+2 \left\langle {\cal D}_3 {\cal D}_1 \right\rangle 
+ \left\langle {\cal D}_2^2 \right\rangle 
+ \left\langle {\cal D}_2 {\cal D}_1^2 \right\rangle ,
\label{eq:AB}
\end{eqnarray}
where ${\cal D}_n = N_f [\partial^n \ln \det M / \partial \mu^n]$
and $\mu \equiv \mu_q a$. \\
${\cal D}_1=N_f {\rm tr} [(\partial M / \partial \mu) M^{-1}],
\hspace{1mm} 
{\cal D}_2=N_f[{\rm tr} [(\partial^2 M / \partial \mu^2) M^{-1}]
 - {\rm tr} [(\partial M / \partial \mu)
              M^{-1} (\partial M / \partial \mu) M^{-1}]],
\ldots $ \\
The derivative of the fermion matrix $M$ at $\mu=0$ is 
\begin{eqnarray}
\frac{\partial^{n} M}{\partial \mu^{n}} = \left\{ 
\begin{array}{l}
-K \left(
 (1-\gamma_4) U_4(x)~\delta_{x+\hat{4},y} 
 -(1+\gamma_4) {U_4}^\dagger(x-\hat{4})~\delta_{x-\hat{4},y} \right)
\ \ {\rm for} \ n: {\rm odd.}
 \\
-K \left(
 (1-\gamma_4) U_4(x)~\delta_{x+\hat{4},y} 
 +(1+\gamma_4) {U_4}^\dagger(x-\hat{4})~\delta_{x-\hat{4},y} \right) 
\ \ {\rm for} \ n: {\rm even.}
\end{array}
\right. 
\end{eqnarray}

For the calculation of these operators, the random noise method is used.
We generate $N_{noise}$ of U(1) noise vectors 
$\left( \eta_{i, \alpha} \right)_{x, \beta} \equiv 
\eta(i,x) \delta_{\alpha, \beta}$, where $\eta(i,x)$ 
is a U(1) random number 
$(\eta=e^{i \theta}; 0 \leq \theta < 2 \pi)$ which satisfies 
$(1/N)\sum_{i=1}^{N} \eta(i,x) \eta^*(i,y)=\delta_{x,y}$
for large $N$. 
$\alpha = 1, \cdots, 12$ is the color and spinor index. 
Then
$\lim_{N \to \infty} (1/N) \sum_{i=1}^{N} \sum_{\alpha=1}^{12} 
\left( \eta_{i, \alpha} \right)_{x, \beta}
\left( \eta_{i, \alpha}^* \right)_{y, \gamma}
=\delta_{x,y} \delta_{\beta, \gamma}$, hence
\begin{eqnarray}
{\rm tr} \left( \frac{\partial^n M}{\partial \mu^n} M^{-1}
 \cdots M^{-1} \right)
& \approx & \frac{1}{N_{noise}} \sum_{i=1}^{N_{noise}} \sum_{\alpha=1}^{12}
\eta_{i, \alpha}^{\dagger} 
\frac{\partial^n M}{\partial \mu^n} X_{i, \alpha}, 
\hspace{1cm} (n=1, 2, \cdots),
\label{eq:noise}
\end{eqnarray}
where $X_{i, \alpha}$ is the solution of 
$MX_{i, \alpha}=(\cdots M^{-1})\eta_{i,\alpha}$.

\section{Quark number and isospin susceptibilities}

\begin{figure}[t]
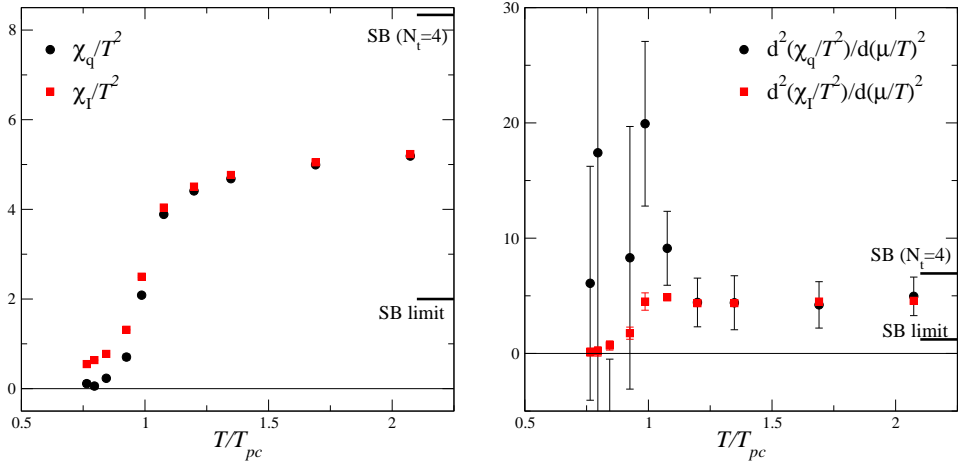

\begin{center}
\includegraphics[width=2.4in]{qnsvTno1050c2.eps}
\hskip 0.5cm
\includegraphics[width=2.4in]{qnsvTno1050c4.eps}
\vskip -0.2cm
\caption{Left: Quark number (black) and isospin (red) susceptibilities.
Right: The second derivatives of these susceptibilities.}
\label{fig2}
\end{center}
\vskip -0.3cm
\end{figure} 

We calculate the expansion coefficients $c_2, c_4, c_2^I$ and $c_4^I$.
The black and red symbols in Fig.~\ref{fig2} (left) are the quark 
number susceptibility $\chi_q/T^2=2c_2$ and isospin susceptibility 
$\chi_I/T^2=2c_2^I$, respectively.
We also plot preliminary results of the second derivatives of 
these susceptibilities 
$\partial^2 (\chi_q/T^2)/ \partial (\mu_q/T)^2 = 24 c_4$ (black symbols) 
and $\partial^2 (\chi_I/T^2)/ \partial (\mu_q/T)^2 = 24 c_4^I$ 
(red symbols) in Fig.~\ref{fig2} (right).
In this study, we choose $N_{noise}=10$ for the calculations of 
the operators in Eq.~(\ref{eq:AB}) except for the operators 
${\rm tr}[(\partial^n M/\partial \mu^n) M^{-1}]$, where $n=1-4$. 
We increase the number of noise vectors up to $N_{noise}=50$ 
for ${\rm tr}[(\partial^n M/\partial \mu^n) M^{-1}]$
to efficiently reduce statistical errors. (See discussions below.)

To check the reliability of the random noise method, we calculate 
the operators ${\cal D}_n$ $(n=1 - 4)$ using two independent sets of 
noise vectors with $N_{noise}=10$ on the same configurations .
Figure \ref{fig3} (left) shows the time history of the imaginary part 
of ${\cal D}_1$ and the real part of ${\cal D}_2$ computed 
by two series of noise sets.
The operator ${\cal D}_n$ is real for $n$ even and purely imaginary 
for $n$ odd, and the expectation value of ${\cal D}_1$ is zero because 
an expectation value of imaginary part is always zero at $\mu_q=0$ 
\cite{BS02}.
It is found that two results of ${\cal D}_2$ obtained by different 
noise sets are consistent with each other on each configuration, 
while two results of ${\cal D}_1$ are sensibly different. 
This means that errors from the noise method is dominating in ${\cal D}_1$
with $N_{noise}=10$. Moreover we found that the error from ${\cal D}_1$
dominates in $c_4$ and $c_4^I$ through Eq.~(\ref{eq:AB}). Therefore, in reducing
the errors for second derivatives, it is efficient to increase
$N_{noise}$ for ${\cal D}_1$, keeping $N_{noise}=10$ for other operators.

As seen in Fig.~\ref{fig2} (left), $\chi_q/T^2$ and $\chi_I/T^2$ 
increase sharply at $T_{pc}$, in accordance with the expectation 
that the fluctuations in the quark-gluon plasma phase are much larger 
than those in the hadron phase.
These results agree qualitatively with previous results by 
staggered-type quarks \cite{BS03,Gottlieb}. 
Moreover the result of $\partial^2 (\chi_I/T^2)/ \partial (\mu_q/T)^2$ is 
quite similar to the results by p4-improved staggered fermions \cite{BS03}. 
This suggests that there are no singularities in $\chi_I$ at 
non-zero density, as discussed in Ref.~\cite{BS03}. 
On the other hand, we expect a large enhancement in the quark number 
fluctuations near $T_{pc}$ as approaching the critical endpoint in the 
$(T, \mu_q)$ plane. 
Although the results in Fig.~\ref{fig2} (right) have large statistical 
errors, the results of $\partial^2 (\chi_q/T^2)/ \partial (\mu_q/T)^2$ 
near $T_{pc}$ show quite different behavior from those of $\chi_I$. 
The second derivative of $\chi_q$ near $T_{pc}$ seems to be more 
than three times larger than those at high temperature, 
suggesting the large fluctuations near the critical point.
However, the lattice discretization error in the equation of state is 
known to be large for the action in Eq.~(\ref{eq:fermact}).
The short lines in the right end denote the values in the free 
quark-gluon gas (Stefan-Boltzmann) limit, both for $N_t=4$ and 
in the continuum. 
It is clear that we need further studies increasing the statistics 
and decreasing the lattice spacing.

\section{Lines of constant pressure} 

\begin{figure}[t]
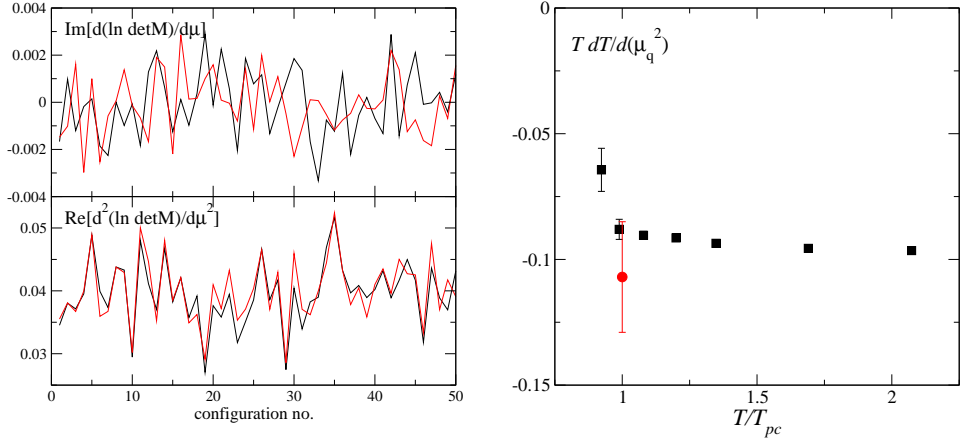

\begin{center}
\includegraphics[width=2.4in]{thd1d2_b180.eps}
\hskip 0.5cm
\includegraphics[width=2.4in]{tdtdmu.eps}
\vskip -0.2cm
\caption{Left: Time history of ${\cal D}_1$ (top) and ${\cal D}_2$ 
(bottom) obtained by different noise sets at 
$T/T_{pc}=0.925, m_{PS}/m_V=0.8$.
Right: Slope of the line of constant pressure at $\mu_q=0$.}
\label{fig3}
\end{center}
\vskip -0.3cm
\end{figure} 

It is interesting to compare the line of constant pressure 
(and energy density) with $T_{pc}(\mu_q)$ and the chemical freeze out 
points \cite{Braum}.
Here we consider the pressure constant line in the $(T, \mu_q^2)$ 
plane, 
\begin{eqnarray}
\Delta p = \frac{\partial p}{\partial T} \Delta T 
+ \frac{\partial p}{\partial (\mu_q^2)} \Delta (\mu_q^2) 
= \left[ T^4 \frac{\partial (p/T^4)}{\partial T} + \frac{4p}{T} \right] 
\Delta T 
+ \left[ T^4 \frac{\partial (p/T^4)}{\partial (\mu_q^2)} \right]
\Delta (\mu_q^2) =0. 
\end{eqnarray}
From this equation, the slope of the $p$-constant line at $\mu_q=0$ is 
given by
\begin{eqnarray}
T\frac{{\rm d} T}{{\rm d} (\mu_q/T)^2} = \left. 
- \frac{\partial(p/T^4)}{\partial (\mu_q/T)^2} 
\right/ \left( T\frac{\partial (p/T^4)}{\partial T} + \frac{4p}{T^4} \right).
\end{eqnarray}
Using the data of $\partial(p/T^4)/\partial (\mu_q/T)^2 = \chi_q/T^2$ 
in Fig.~\ref{fig2}, $p/T^4$ and $T\partial (p/T^4) / \partial T$ 
in Ref.~\cite{EoS},
we estimate the slope of the line of constant pressure at $\mu_q=0$.
The results at $m_{PS}/m_V=0.8$ are shown in Fig.~\ref{fig3} (right).
The slope at $\mu_q=0$ is about $-0.1$. This is roughly consistent with 
the previous results by p4-improved staggered fermions at 
$m_{PS}/m_V=0.7$ in Ref.~\cite{BS02}, which is denoted by the red dot.
Moreover the slope of the phase transition line in Ref.~\cite{BS02} is 
$T {\rm d} T_{pc} / {\rm d} (\mu_q/T)^2 \approx -0.07(3)$. 
The line of constant $p$ is almost parallel to the phase transition line.
On the other hand, the slope $T {\rm d} T / {\rm d} (\mu_q/T)^2$ of 
the line of the chemical freeze out in the $(T, \mu_q^2)$ plane are 
about $-0.25$ \cite{Braum}.
Further studies at small quark mass and large $N_t$ are necessary 
to compare with experimental results.

\section{Conclusion}
We reported the current status of our study of QCD thermodynamics 
with a Wilson-type quark action. 
The lines of constant physics in the $(\beta,K)$ plane as well as 
the relation between the parameters $(\beta,K)$ and 
$(T/T_{pc}, m_{PS}/m_V)$ are determined.
Simulations are performed on a $16^3 \times 4$ lattice.
The derivatives of pressure with respect to $\mu_q$ and $\mu_I$ 
up to 4th order are computed.
The random noise method is used. For the calculation of 4th order 
derivatives, the choice of the number of noise vector $(N_{noise})$ 
is important.
We discussed the fluctuations of quark number and isospin densities.
Although the statistical errors are still large, a clear quantitative 
difference between the second derivatives of $\chi_q$ and $\chi_I$ 
is observed. 
$\chi_q$ seems to increase rapidly near $T_{pc}$ as $\mu_q$ increases, 
whereas the increase of $\chi_I$ is not large near $T_{pc}$. 
These behaviors agree with the results obtained by p4-improved 
staggered fermions qualitatively, 
and with the expectation from the sigma model.

\paragraph{Acknowledgements:}
This work is in part supported by Grants-in-Aid of the Japanese 
 Ministry of Education, Culture, Sports, Science and Technology, 
(Nos.~13135204, 15540251, 17340066, 18540253, 18740134).
SE is supported by the Sumitomo Foundation (No.~050408), and 
YM is supported by JSPS.
This work is in part supported also by the Large-Scale Numerical
Simulation Projects of ACCC, Univ. of Tsukuba, and by the Large Scal
Simulation Program of High Energy Accelerator Research Organization
(KEK).

\end{document}